\documentclass[modern]{aastex701}

\usepackage{graphicx} 
\usepackage[margin=1in, letterpaper]{geometry} 

\usepackage{amsmath}
\usepackage{soul}

\newcommand{\afe}[0]{[\alpha/{\rm Fe}]}

\newcommand{\feh}[0]{[{\rm Fe/H}]} 
\newcommand{\xfe}{[{\rm X}/{\rm Fe}]}

\newcommand{\logg}{\log(g)}
\newcommand{\teff}{T_{\rm eff}}

\newcommand{\kel}{\rm \; K}
\newcommand{\msun}{M_{\odot}}
\newcommand{\sphottwo}{\sigma_{\rm phot, \,2-param}}

\newcommand{\sphotmed}{\sigma_{\rm phot, \,med}}

\submitjournal{ApJ}

\shorttitle{Abundance Scatter}
\shortauthors{Griffith and Blum}
\renewcommand{\paragraph}[1]{\bigskip\par\noindent{\textbf{#1}}~---}
\sloppypar

\graphicspath{{./}}

\begin{document}

\title{Untangling the Sources of Abundance Dispersion in Low-metallicity Stars II: Neutron Capture Elements}

\author[0000-0001-9345-9977]{Emily J. Griffith}
\altaffiliation{Hubble Fellow}
\affiliation{Center for Astrophysics and Space Astronomy, Department of Astrophysical and Planetary Sciences, University  of Colorado, 389~UCB, Boulder,~CO 80309-0389, USA}
\email{Emily.Griffith-1@colorado.edu}

\author{Marissa Blum}
\affiliation{Center for Astrophysics and Space Astronomy, Department of Astrophysical and Planetary Sciences, University  of Colorado, 389~UCB, Boulder,~CO 80309-0389, USA}
\email{Marissa.Blum@colorado.edu}

\author[0000-0001-7775-7261]{David H. Weinberg}
\affiliation{The Department of Astronomy and Center of Cosmology and AstroParticle Physics, The Ohio State University, Columbus, OH 43210, USA}
\affiliation{The Institute for Advanced Study, Princeton, NJ, 08540, USA}
\email{weinberg.21@osu.edu}

\author[0000-0001-7258-1834]{Jennifer A. Johnson}
\affiliation{The Department of Astronomy and Center of Cosmology and AstroParticle Physics, The Ohio State University, Columbus, OH 43210, USA}
\email{johnson.3064@osu.edu}

\author[0000-0001-8208-9755]{Tawny Sit}
\affiliation{The Department of Astronomy and Center of Cosmology and AstroParticle Physics, The Ohio State University, Columbus, OH 43210, USA}
\email{sit.6@osu.edu}

\author{Ilya Ilyin}
\affiliation{Leibniz-Institut for Astrophysics Potsdam (AIP), An der Sternwarte 16, D14482 Potsdam, Germany}
\email{ilyin@aip.de}

\author[0000-0002-6192-6494]{Klaus G. Strassmeier}
\affiliation{Leibniz-Institut for Astrophysics Potsdam (AIP), An der Sternwarte 16, D14482 Potsdam, Germany}
\email{kstrassmeier@aip.de}

\begin{abstract}\noindent 
We present the abundances of 23 elements, including 11 heavy elements (Cu, Zn, Sr, Y, Zr, Ba, La, Ce, Nd, Sm, Eu) for up to 86 metal-poor ($-2 \lesssim \feh \lesssim -1$) subgiants. 
We use KORG, a state of the art spectral synthesis package, to derive 1D-LTE abundances from high-SNR and high-resolution spectra taken by the Large Binocular Telescope with the Potsdam Echelle Polarimetric and Spectroscopic Instrument. These precise spectra and abundance measurements minimize the impact of photon-noise ($\lesssim0.06$ dex), allowing us to robustly measure the intrinsic abundance scatter in [X/Fe]. After removing two stars with exceptional $s$-process enhancement, we find that the intrinsic scatter among the $s$- and $r$-process elements tends to be larger than for the lighter elements, with heavy element scatter ranging from 0.11 (Zn) to 0.27 (Eu) dex. Intrinsic abundance scatter could have multiple origins, including star-to-star variations in the ratios of nucleosynthetic sources as well as stochastic sampling of the progenitor supernovae properties, such as mass, rotation, and magnetic field strength. We explore the expected abundance scatter signature caused by stochastic sampling, finding that a fraction of both rapidly rotating CCSN and magnetorotationally driven SN are needed to reach the observed abundances and intrinsic scatter. This analysis is limited by the restrictive parameter spaces spanned by existing yield sets. A diverse, finely sampled grid of supernovae yields is needed to robustly model stochastic abundance scatter.

\end{abstract}


\section{Introduction}\label{sec:intro}

The abundance spreads observed in the low-metallicity Milky Way (MW) disk, halo, and local dwarf galaxies are well matched by stochastic models of Galactic chemical evolution \citep{cescutti2008, cescutti2014, alexander2023, lucertini2025} with enrichment from finite numbers of supernova (SN) events and a stochastically sampled initial mass function (IMF) \citep[e.g.,][]{carigi2008}.
These comparisons can be powerful tools to constrain the MW's detailed enrichment history---with heavy element abundance scatter probing the uncertain nucleosynthetic channels that produce the neutron capture elements \citep[e.g.,][]{francois2007,brauer2021}
To date, however, few large samples of stellar abundance derived from high-resolution ($R\gtrsim 50,000$), high-signal-to-noise (S/N$\gtrsim100$) spectra exist, limiting our ability to constrain the scatter signature in the low-metallicity MW disk and halo. In this paper, we present such abundances and their intrinsic abundance scatter for a sample of nearby subgiants of $-2\leq\feh\leq-1$, interpreting our results through a lens of stochastic enrichment. 

Unlike mean abundance trends, which inform us about star formation history, gas accretion, and relative contributions of nucleosynthetic sources \citep[e.g.,][]{tinsley1980, matteucci1986, mcwilliam1997, andrews2017, weinberg2017, spitoni2019, weinberg2019}, abundance scatter is a probe of stochastic enrichment from the SN themselves and star-to-star variations in the relative nucleosynthetic source fractions \citep{cescutti2014,griffith2023}. 
Past works including \citet{bertran2016} and \citet{vincenzo2021} quantify scatter in [$\alpha$/Fe] for the high-Ia (low-$\alpha$) and low-Ia (high-$\alpha$) disk, finding typical intrinsic scatter of $\sim0.04$ dex at $-0.7 \lesssim\feh\lesssim0.4$.
Here, abundance scatter is predominantly driven by star-to-star variations in the ratio of core-collapse supernovae (CCSN), Type-Ia supernovae (SNIa), and asymptotic giant branch (AGB) stars \citep[e.g.,][]{weinberg2019, griffith2019, griffith2024, sit2025}. When this source of scatter is removed, the remaining intrinsic scatter signature is $\leq 0.02$ dex for most elements \citep{ratcliffe2022, ting2022, griffith2025}.

At very low metallicity ($-4 \lesssim\feh\lesssim-2$), the observed abundance scatter in the MW halo is much larger, ranging from 0.05 to 0.2 dex for $\alpha$ and Fe-peak elements \citep[e.g,][]{cayrel2004, lombardo2022, Li2022} and drastically increasing to 0.5 dex for many neutron capture elements \citep[e.g.,][]{francois2007, lombardo2025}. In this low-metallicity regime, scatter in certain abundance ratios, such as [Sr/Ba] or [Eu/Ba], is used to motivate the need for multiple enrichment sources, and correlations in abundance scatter between elements can constrain like nucleosynthetic channels \citep[e.g.,][]{travaglio2004, spite2018, lombardo2025}.

To fill the gap between these two metallicity spaces and measure \textit{intrinsic} abundance scatter in the MW, \citep[][hereafter G23]{griffith2023} present abundances of 12 light elements for a sample of 86 subgiants stars with $-2\lesssim\feh\lesssim-1$---homogeneously observed and analyzed to minimize abundance systematics. With a high-resolution boutique study, they achieve higher abundance precision and more robust scatter estimates than similar works with medium resolution spectroscopic surveys \citep[e.g.,][]{naidu2020, belokurov2022, horta2022}. By comparing the measured intrinsic scatter in $\alpha$ and Fe-peak elements to an estimate of the scatter signature expected for stochastic sampling of the IMF, G23 find that a typical star in their sample has likely been enriched by $\sim50$ CCSN, corresponding to a gas mixing mass of $\sim 6\times10^4\msun$. 

In this paper, we extend the analysis of G23 to the heavier elements, focusing on those produced by the slow neutron capture process ($s$-process) and the rapid neutron capture process ($r$-process). Unlike the light elements, which are predominantly produced by CCSN and SNIa, the heavy elements are produced through a diverse set of enrichment channels capable of neutron capture \citep{cowan2021}. The $s$-process elements (e.g., Sr, Y, Ba) are dominantly produced in rotating CCSN \citep[e.g.,][]{limongi2018, choplin2018, prantzos2018} and AGB star winds \citep[e.g.,][]{karakas2010, cristallo2011}. While non-rotating massive stars cannot produce such species, rotational mixing between the H-shell and the He-core in massive stars rotating at $\gtrsim40\%$ of critical velocity produce primary $^{14}\text{N}$ and $^{22}$Ne, boosting the production the free neutrons needed for the $s$-process \citep{frischknecht2016}. Rotating massive stars can produce $s$-process nuclei up to Ba, with yields strongly dependent upon the the initial rotation rate and the initial metallicity, which sets the Fe seed to free neutron ratio. Because massive stars are short lived, rotating CCSN contribute promptly and are a large production source of light $s$-process nuclei at low metallicity. Enrichment from finite numbers of CCSN with varying masses and rotation rates may cause abundance scatter.

Lower-mass AGB stars also produce $s$-process material through a complex interplay of shell mixing and thermal pulses. Unlike rotating CCSN, AGB stars can produce both light (e.g. Sr, Zr) and heavy (e.g. Ba, Ce) $s$-process material \citep{karakas2014}. Though AGB stars are longer lived than CCSN and have strongly metallicity dependent yields, the $s$-process yields are largest for AGB stars with $M\approx3\msun$ and $\sim500$ Myrs lifetimes \citep[e.g.,][for Sr]{johnson2020}. This suggests that though AGB stars contribute most heavily to nucleosynthesis at $\feh\geq-1$, they produce non-negligible quantities of heavy elements at lower metallicities. In their AGB only GCE models, \citet{kobayashi2020} find [X/Fe] $\gtrsim -2$ at [Fe/H] $> -2$ for most $s$-process elements, though these models cannot reach solar [X/Fe] abundance ratios at [Fe/H] $< -1$. 

The main, low-metallicity $r$-process source is far less constrained. While the GW170817 neutron star merger event confirmed that some of the heaviest elements are produced in binary neutron star mergers \citep{abbott2017}, it is not yet known if these events can produce sufficient $r$-process yields to explain low-metallicity stellar abundances \citep[e.g.,][]{argast2004, kobayashi2020}. Other extreme, rare SN have been proposed as additional astrophysical sites for the $r$-process, including magnetorotationally driven SN \citep[CCSN powered by strong magnetic fields around the proto-neutron star which launch jets; e.g.,][]{duncan1992,nishimura2006,nishimura2015}, electron capture SN \citep[collapse of an $8-10\msun$ star from electron capture; e.g.,][]{wanajo2009,wanajo2018}, and collapsars \citep[collapse of a rapidly rotating massive star which induces accretion onto the black hole remnant and relativistic jets; e.g.,][]{siegel2019, issa2025}. Though the yields of these sources are uncertain, their rare nature likely results in observable abundance dispersion \citep[e.g.,][for Eu in collapsars]{brauer2021}.

The abundance scatter observed at low-metallicity may help constrain the relative contribution of these nucleosynthetic sources to the production of $s$- and $r$-process material in the early MW. Such measurements require high-resolution and high-S/N observations to ensure that the intrinsic scatter can be robustly measured against the background photon-noise. While studies including \citet{nissen2011} and the Measuring at Intermediate Metallicity Neutron Capture Elements survey \citep[MINCE;][]{cescutti2022, francois2024, lucertini2025} measure heavy element abundances from high-resolution spectra at $-2 \leq \feh \leq-1$, neither are suitable for this analysis. The \citet{nissen2011} sample of 94 stars with $-1.6 \leq \feh \leq -0.4$ spans the $\alpha$ and Fe-peak elements, but only includes two $s$-process elements (Ba and Y). While the MINCE sample of $65$ stars with $-2.5 \leq \feh \leq -1.5$ includes ten neutron capture elements, their sample is observed by many facilities with unique systemics---introducing a source of abundance dispersion beyond photon-noise. 

The observations and abundance analysis in this work were optimized to minimize abundances systematics, allowing us to place robust constraints on the intrinsic scatter of over 20 elements, including nine produced by neutron capture. We summarize our observations and data reduction in Section~\ref{sec:data}. In Section~\ref{sec:analysis}, we describe our abundance determination methods, leveraging new spectral synthesis code \texttt{Korg} \citep{wheeler2023,wheeler2024}. We compare light element abundances derived here with those from G23 and heavy element abundances with GALAH and the MINCE survey. We present our intrinsic scatter measurements in Section~\ref{sec:intrin-scatter} and discuss their implications for stochastic sampling of the underlying SN population in Section~\ref{sec:disc}. Finally, we summarize our conclusions and discuss the limitations of existing yield sets in Section~\ref{sec:summary}

\section{Observations and Data Reduction}\label{sec:data}

In this work we adopt the stellar sample curated by G23. 
This sample consists of 102 metal-poor ($-2 \leq [\rm Fe/H] \leq -1$), bright ($V < 12.5$), nearby ($d < 2500$ pc) subgiants ($3.3 < \logg < 4$; $2 < G < 4.3$; $0.75 < G_{BP} - G_{RP} <1.1$), selected based on stellar parameters from the Large sky Area Multi-Object fiber Spectroscopic Telescope (LAMOST) DR5 \citep{xiang2019} and Gaia eDR3 magnitudes \citep{gaia2021}, with distances and absolute $V$-band magnitudes calculated with the astroNN Gaia tools\footnote{\url{https://astronn.readthedocs.io/en/latest/tools_gaia.html}} \citep{bovy2017}. 
G23 chose these cuts to maximize the number of observable targets while achieving light element abundance precision of $\leq 0.04$ dex and minimizing the effects of systematic abundance uncertainties due to abundance correlations with $\logg$ and $\teff$ \citep{holtzman2018, jofre2019, griffith2021a}. 

Of the 102 targets which meet the sample criteria, G23 obtain high-resolution optical spectra of 98 stars. Spectra were taken with the Potsdam Echelle Polarimetric and Spectroscopic Instrument \citep[PEPSI;][]{strassmeier2015} on the Large Binocular Telescope (LBT) between 2021 April 30 and 2022 February 12 using the 300 $\mu$m fiber and cross-dispersers (CD) II and IV. This corresponds to $R=\lambda/\Delta\lambda=50,000$ spectra for wavelength ranges of $4260-4800$ and $5440-6279$ $\AA$. Observing with the largest fiber diameter available on PEPSI allows for S/N $>100$ to be reached in both the red and blue channels with exposure times of $300-1200\,s$. CDs II and IV were chosen to maximize the number of atomic species present in the spectral ranges. Median S/N of 125 and 236 are achieved in CD II and IV, respectively. Details of the individual observations are provided in Table 1 of G23.

Spectral reduction is described in detail in G23. In short, spectra are first reduced by the Spectroscopic Data Systems in the PEPSI pipeline \citep{strassmeier2018}. The pipeline applies bias and flat-field corrections, estimates photon-noise, and implements scattered light subtraction. It defines, extracts, and merges the spectral orders, calibrates the wavelength scale against Th-Ar lamp lines, and applies a global continuum correction. G23 continue spectral reduction with the integrated spectroscopic framework, iSpec \citep{blanco2014}. Here, spectra are shifted to the rest frame, telluric lines are masked, and a more detailed continuum normalization is applied.

\section{Stellar Abundance Determination}\label{sec:analysis}

In this work we re-derive the 1D LTE abundances of 12 light elements, originally reported in G23, and present abundances for an additional 11 heavier elements---Cu, Zn, Sr, Y, Zr, Ba, La, Ce, Nd, Sm, and Eu. 

\subsection{Line List}\label{subsec:line_list}

We curate a set of spectral features optimized for our low-metallicity spectra, building off of that used in G23 and referencing line lists from \citet{fulbright2000}, \citet{francois2024}, the $R$-Process Alliance \citep{roederer2018, shah2024}, GALAH Data Release 3 \citep[DR3;][]{buder2021}, Gaia-ESO \citep{gilmore2012, heiter2015, heiter2021}, and the solar spectrum \citep{moore1966}. We construct windows of $\pm 3 \text{\AA}$ around each feature. We refine our window list after a preliminary analysis, removing individual windows that are best fit by an abundance that strongly deviates from the global fit or that contain lines too weak to measure at low metallicity. Within each window, we adopt the Gaia-ESO line list \citep{heiter2015, heiter2021}, which includes line parameters such as oscillator strength ($\log(gf)$)\footnote{Note that we do not replace the $\log(gf)$ values for the Mg lines as done in G23, as it is not the focus of this work.}, excitation potential, and hyper-fine structure (HFS) for Sc, V, Mn, Co, Cu, Ba, La, Nd, Sm, and Eu. We list the included species and number of features in Table~\ref{tab:lines}. 

\begin{figure*}
    \centering
    \includegraphics[width=1\linewidth]{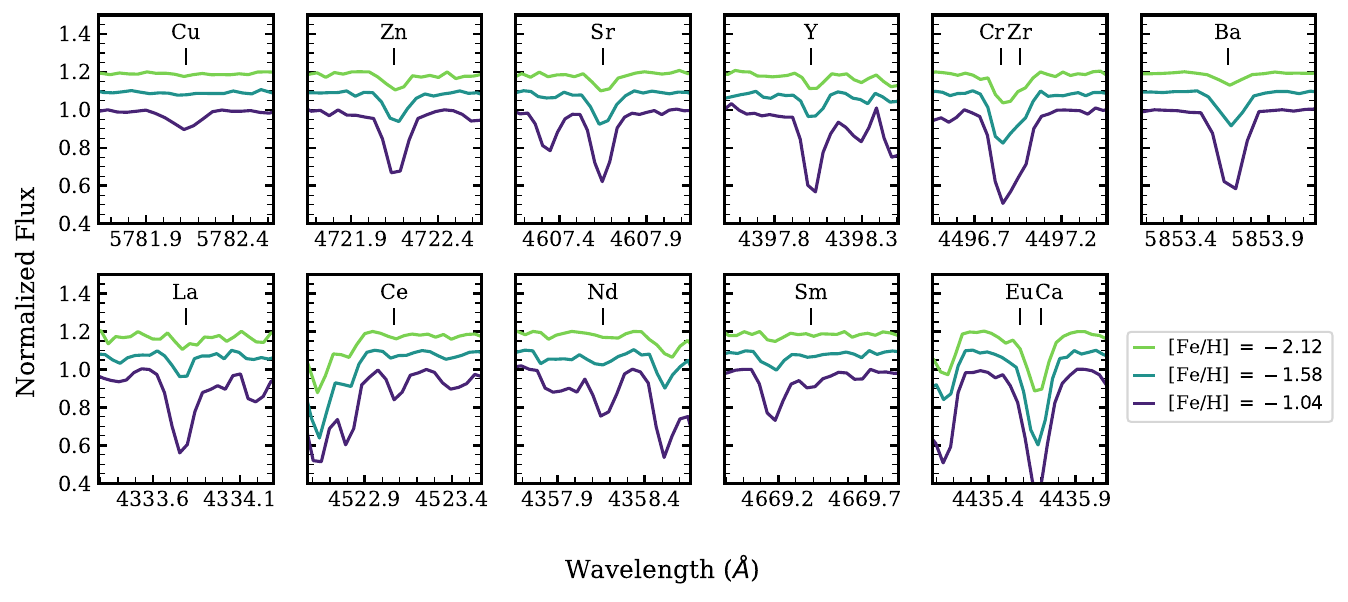}
    \caption{Heavy element line windows for three stars with [Fe/H] of $-2.12$ (green, 2MASS J04315411-0632100), $-1.58$ (blue, 2MASS J15581861+0203059), and $-1.04$ (purple, 2MASS J17140534+1407170). Spectra have been offset by 0.1 for clarity. We show one window for each heavy element, with line centers labeled, including blended lines for Zr and Eu.}
    \label{fig:spectra}
\end{figure*}

In Figure~\ref{fig:spectra}, we plot example line windows for each heavy element presented in this work. In each window, we show spectra for three stars at a range of metallicities that span our parameter space. In this figure, it is apparent that some elements, including Zn, Sr, Y, Zr, Ba, La, Ce, and Nd, have visible lines down to [Fe/H] $\approx -2$, while Cu, Sm, and Eu lines are very weak at low metallicity.

\begin{deluxetable*}{ccc}
\tablecaption{Spectral Features \label{tab:lines}}
    \tablehead{\colhead{Species} &  \colhead{Number of Features}& \colhead{HFS}}
    \startdata
     Na I&  4& No\\
     Mg I&  5& No\\
     Si I&  19& No\\
     Ca I&  26& No\\
     Sc II&  13&  Yes\\
     Ti I&  44&  No\\
     Ti II&  37&  No\\
     V I&  14&  Yes\\
     Cr I&  19&  No\\
     Cr II&  6&  No\\
     Mn I&  16&  Yes\\
     Fe I&  174&  No\\
     Fe II&  23&  No\\
     Co I&  1&  Yes\\
     Ni I&  28&  No\\
     Cu I&  1&  Yes\\
     Zn I&  2&  No\\
     Sr I&  1&  No\\
     Y II&  5&  No\\
     Zr II&  5&  No\\
     Ba II&  3&  Yes\\
     La II&  4&  Yes\\
     Ce II&  10&  No\\
     Nd II&  11&  Yes\\
     Sm II&  11&  Yes\\
     Eu II&  1&  Yes\\
    \enddata
    \tablecomments{Number of spectral features per species included in our analysis. The final column indicates if hyper-fine structure is included for a given species.}
\end{deluxetable*}

\subsection{Spectral Synthesis with \texttt{Korg}}\label{subsec:synthesis}

To derive abundances, we fit the stellar spectra with \texttt{Korg}, a 1D LTE spectral synthesis package \citep{wheeler2023, wheeler2024}. \texttt{Korg}'s abundance fitting routine (\texttt{fit\_spectrum}) seamlessly interpolates MARCS model atmospheres \citep{gustafsson2008}, is capable of fitting complex line profiles, and performs significantly faster than other spectral synthesis codes (e.g., MOOG, used in G23). 

When fitting for individual abundances, we fix parameters $\teff$, $\logg$, [M/H], microturbulent velocity ($v_{\rm micro}$), rotational velocity ($v\sin(i)$), limb-darkening coefficient ($\epsilon$) and spectral resolution ($R$). For consistency with our previous work, we adopt the stellar parameters for $\teff$, $\logg$, [M/H], and $v_{\rm micro}$ derived from MOOG Fe line equivalent width analysis in G23. These values place the Fe I and Fe II lines in ionization and excitation equilibrium. Re-deriving these parameters with \texttt{Korg}'s synthesis routine leads to larger abundance deviations from G23 \citep[e.g.,][]{jofre2019}, but it does not significantly change our abundance scatter results. As in G23, we set $\epsilon=0.6$ and $R=50,000$, the instrument resolution.

We calculate the remaining stellar parameter, $v\sin(i)$, using the \texttt{fit\_spectrum} routine in \texttt{Korg}. Here, $v\sin(i)$ functions as a broadening term encapsulating both rotational velocity and macroturbulent velocity ($v_{\rm macro}$), as \texttt{Korg} is unable to fit for $v_{\rm macro}$ and the two broadening terms may be indistinguishable in our spectra. We fit $v\sin(i)$ using only the Fe I and Fe II lines. Most stars are best fit with $v\sin(i)<5$ km/s, with the sample having a median rotational velocity of $3.38$ km/s. We report all stellar parameters in Table~\ref{tab:params}.

As in G23, we exclude nine stars from our analysis that have stellar parameters falling outside of our target range (five stars with $\teff > 6000 \kel$ and three stars with $\logg < 3$) or show signs of spectroscopic binarity (one star). We proceed in our abundance determination for the remaining 89 stars. Using the \texttt{fit\_spectrum} routine with MARCS model atmospheres \citep{gustafsson2008} and solar abundances from \citet{asplund2021}, we iteratively fit individual elemental abundances, starting with Fe and proceeding from the lightest to heaviest elements. After an element is fit, its abundance is fixed for the remaining analysis. All windows for a given element are simultaneously fit in the synthesis routine, with \texttt{Korg} minimizing the fit $\chi^2$ across all windows. This routine returns the best-fit [X/H] abundances for all stars. Results do not change if we simultaneously fit all 23 elements, though the computation time is longer. 

For abundance features that are not detected (\texttt{Korg} best-fit value of $[\rm X/H]<-3$), we report an upper limit. Upper limits are derived by finding the abundance value that increases the fit $\chi^2$ by 25 ($5\sigma$) relative to a continuum fit with $[\rm X/H]=-5$ \citep[e.g.,][]{ji2020}. We provide [X/H] abundances for all 89 stars in Table~\ref{tab:params}, with upper limits indicated in the abundance error column. As in G23, we find that three stars have [Fe/H] $ > -0.9$, placing them outside of our target metallicity range. We report abundances for these stars but exclude them from our abundance scatter analysis. 

\begin{figure*}
    \centering
    \includegraphics[width=1\linewidth]{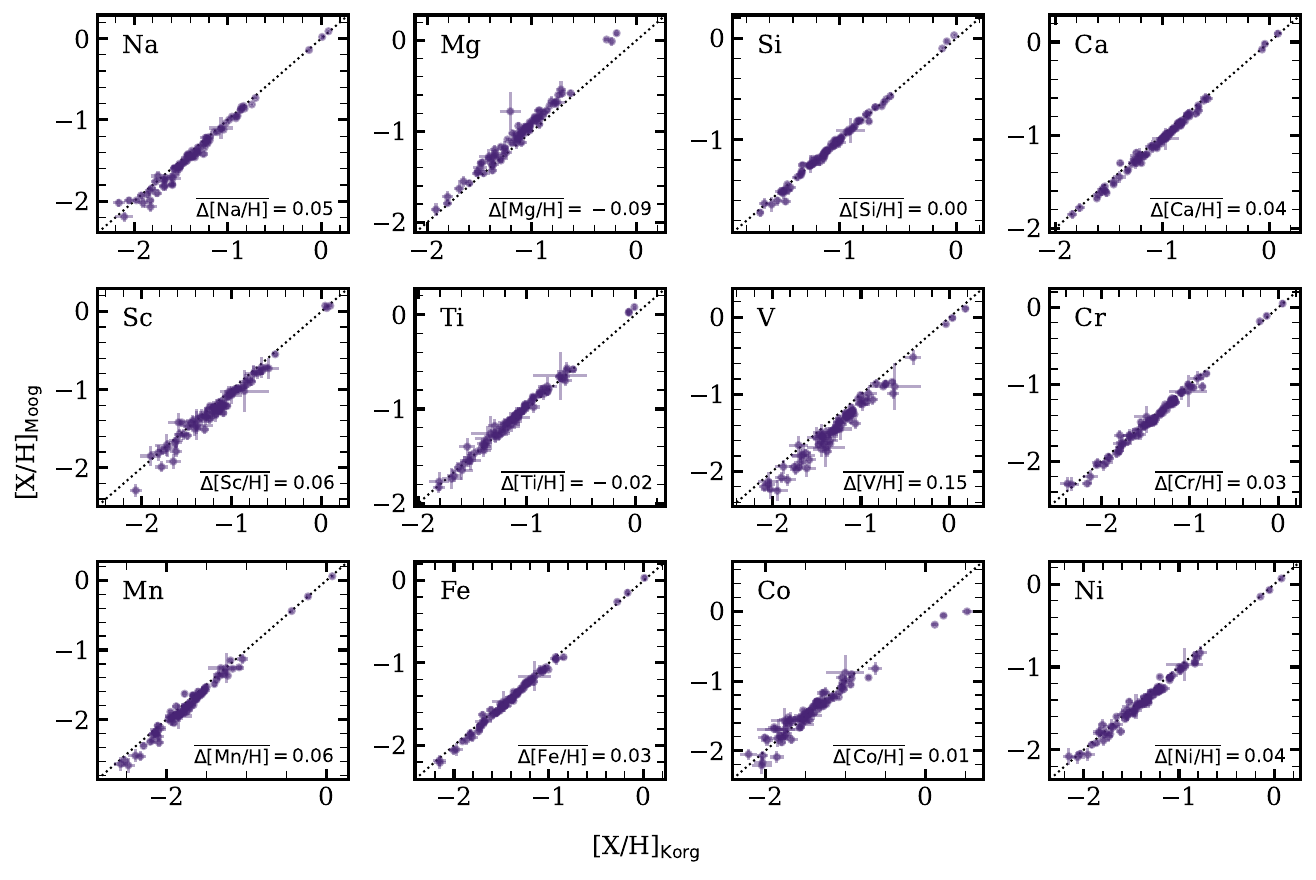}
    \caption{Comparison of [X/H] abundances derived in this work using \texttt{Korg} synthesis and [X/H] abundances derived in G23 using MOOG synthesis. Error bars represent photon-noise estimates on individual abundances, as discussed in Section~\ref{subsec:photon-noise}. In the bottom right corner of each panel we include the average difference in abundances, where $\Delta [\rm X/H]  = [X/H]_{\rm Korg} - [X/H]_{\rm MOOG}$. Note that the axes scales vary between subplots.}
    \label{fig:comp_G23}
\end{figure*}

In Figure~\ref{fig:comp_G23}, we plot the abundances for the light elements (Na through Ni), comparing the values derived through MOOG spectral synthesis in G23 with the values derived through \texttt{Korg} synthesis in this work. We find that the two analysis methods are in strong agreement, with average $\Delta[\rm X/H] \lesssim 0.05$ for most elements. The \texttt{Korg} values derived here tend to be slightly larger than the values from G23. The larger abundance differences for Sc, V, and Mn are not surprising, as we include HFS structure lines in our analysis while these were not included in G23. This comparison shows that \texttt{Korg} performs similarly to other spectral synthesis fitting algorithms and that small changes to the analysis methods or line lists can cause systematic shifts in derived stellar abundances.

\begin{deluxetable*}{cccccccc}
\tablecaption{Stellar Parameters, Abundances, and Uncertainties.  \label{tab:params}}
    \tablehead{\colhead{Object} &  \colhead{[Fe/H]} & \colhead{$\sigma$[Fe/H]}  &  \colhead{[Cu/Fe]} & \colhead{$\sigma$[Cu/Fe]}  &  \colhead{[Zn/Fe]} & \colhead{$\sigma$[Zn/Fe]}   & \colhead{...}}
    \startdata
    2MASS J00021423+3228190 & -1.45 &    0.08 &  -0.18 &       0.06 &   0.23 &       0.03 & ...\\
    2MASS J00091409+1728209 & -1.98 &    0.07 &  -0.43 &       0.12 &   0.13 &       0.04 & ...\\
    2MASS J00383315+3433115 & -1.38 &    0.03 &  -0.52 &       0.05 &   0.17 &       0.03 & ...\\
    2MASS J01051165+3103568 & -1.46 &    0.03 &  -0.34 &       0.10 &   0.19 &       0.04 & ...\\
    2MASS J01591792+0212080 & -1.06 &    0.03 &  -0.17 &       0.04 &   0.24 &       0.02 & ...\\
    \enddata
    \tablecomments{The full table with parameters and abundances for all stars is available online.}
\end{deluxetable*}

\subsection{Photon-Noise}\label{subsec:photon-noise}

By selecting a uniform stellar sample of subgiants and constructing a standardized analysis pipeline, we have minimized systematic abundance errors \citep[e.g.,][]{jofre2019} and assume they have a negligible effect on our results. The main contribution to external scatter on the abundances derived in this work is instead photon-noise. We expect the photo-noise to be small ($\lesssim 0.04$ dex) due to our high-resolution and high-S/N observations. Photon-noise is often quantified through analysis of multiple observations of the same stars \citep[e.g.,][]{jofre2019,hayes2022} or analysis of reference stars with well known abundances \citep[e.g.,][]{hawkins2016}. While we do not have repeat observations of our targets nor spectra of reference stars, we are able to simulate multiple observations using the flux error spectrum, as described in G23. 

To simulate such observations, G23 vary the spectral flux at each wavelength, adding to the observed flux a random fluctuation drawn from a Gaussian distribution with width equal to the flux error at that wavelength. They repeat this spectral variation ten times, creating ten unique spectra for each star, and re-derive stellar parameters for each spectral iteration. This process propagates uncertainty in the spectral flux to the stellar parameters and stellar abundances, accounting for the many ways that photon-noise impacts spectral analysis. We repeat the spectral synthesis analysis described in Section~\ref{subsec:synthesis} with the varied spectra and associated stellar parameters from G23. We quantify the photon-noise on each individual abundance as the standard deviation of the abundances derived from the ten spectra with random variations. We report this as $\sigma_{\rm [X/Fe]}$ in Table~\ref{tab:params}. We find that the median $\sigma_{\rm [X/Fe]}$ on detected abundances is less than or equal to 0.050 dex for all elements but Cu ($\sigma_{\rm [Cu/Fe],\, med}$ = 0.055) and Eu ($\sigma_{\rm [Y/Fe],\, med}$ = 0.071). 

\subsection{Abundance Trends and Comparisons with Previous Works}\label{subsec:abd_trends}

\begin{figure*}
    \centering
    \includegraphics[width=1\linewidth]{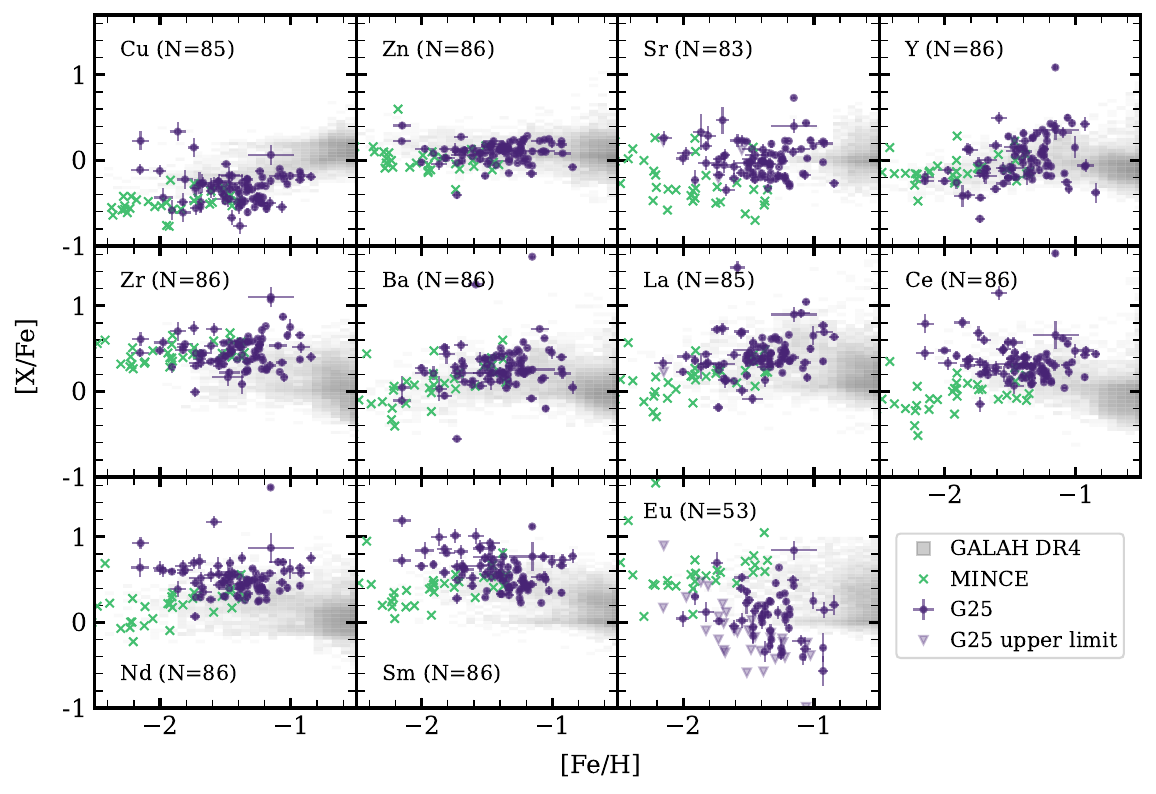}
    \caption{[X/Fe] vs. [Fe/H] abundances for the heavy element abundances derived in this work (dark purple points). Error bars represent the photometric-noise (Section~\ref{subsec:photon-noise}) Stars for which only upper limits can be measured are shown as light purple downward triangles. In each panel, we note the number of stars with a detected abundances of a given element out of the 86 total low-metallicity stars. For comparison, we also include a sample of high S/N stars from GALAH DR4 (grey 2D histogram; \citealp{buder2025}) and low-metallicity stars from the MINCE survey (teal xs; \citealp{cescutti2022, francois2024}).}
    \label{fig:x_fe}
\end{figure*}

We report the stellar abundances and uncertainties of 89 stars (86 low-metallicity) for 23 elements in Table~\ref{tab:params} and show the heavy element [X/Fe] vs. [Fe/H] abundances trends for the low-metallicity stars in Figure~\ref{fig:x_fe}. In this figure, we include a sample of unflagged, high S/N ($>100$) stars from GALAH DR4 \citep{buder2025} as well as the MINCE I and II abundances \citep{cescutti2022, francois2024}. At $\feh < -1$, both samples predominantly consist of evolved giant stars, but provide a comparison for abundances trends at higher and lower metallicities. We generally see good agreement in the abundance trend location and dispersion between our stars and the GALAH and MINCE samples.

The heavy elements discussed here fall into four rough categories: Fe-cliff, first $s$-process peak, second $s$-process peak, and $r$-process. The ``Fe-cliff,'' a label coined by \citet{griffith2019}, consists of Cu and Zn. These elements lie on the steeply falling edge of the Fe-peak and likely have a unique nucleosyntheic origin compared to Fe-peak elements. At low-metallicity, Cu and Zn are likely produced entirely through explosive nucleosynthesis in CCSN \citep[e.g.,][]{andrews2017, rybizki2017}. Significant Cu and Zn yields can be obtained in non-rotating CCSN, but rapid rotation enhances yields and results in predicted [Cu/Fe] and [Zn/Fe] near observations \citep{limongi2018, choplin2018, prantzos2018}. We observe both the Cu and Zn abundance trends to be an extension of the MINCE trends, with visually similar levels of abundance dispersion. The Cu abundances are offset from the GALAH trend to slightly lower [Cu/Fe]. The [Zn/Fe] appears constant with metallicity near the solar value, while the [Cu/Fe] abundances are metallicity dependent and sub-solar for $\feh < -1$. In both Cu and Zn we see outliers to larger [X/Fe] in the lowest metallicity stars. The two stars with large [Zn/Fe] have strong, visible Zn lines in their spectra that are well fit. The five star with [Cu/Fe]$>-0.2$ and [Fe/H]$<-1.5$ have statistically significant fits, suggesting that they are not upper limits, though the Cu line in their spectra is difficult to distinguish from the continuum by eye.

The first $s$-process peak elements consist of Sr, Y, and Zr. These elements sit near magic neutron number of 50, with 65\% to 75\% of their solar abundance produced by the $s$-process \citep{bisterzo2014}. At low metallicity, these elements are thought to be dominantly produced by rapidly rotating CCSN \citep{limongi2018, choplin2018}, though AGB star winds and rare $r$-process events may also contribute to their production \citep[e.g.,][]{cristallo2011,nishimura2015, wanajo2018}. We observe the [Sr/Fe], [Y/Fe], and [Zr/Fe] abundances to be predominantly super solar. The total scatter and photon-noise scatter appears larger than for the Fe-cliff elements. [Y/Fe] shows a strong metallicity dependence, with the most metal rich end of our sample ($\feh \approx -1$) beginning to turn over towards the solar abundance. The [Sr/Fe] and [Y/Fe] trends appear as an extension of the GALAH trends, while the [Zr/Fe] is offset to larger values than the GALAH sample. We find good agreement with the MINCE abundance trends, though their [Sr/Fe] appears more scattered and centered at a lower abundance than our values. 

Ba, La, Ce, and Nd fall in the second $s$-process peak, found near a magic neutron number of 82. While only 57\% of the solar Nd abundance is produce by the $s$-process, it is responsible for $65-75\%$ of the solar Ba, La, and Ce \citep{bisterzo2014}. 
Unlike the elements in the first $s$-process peak, those in the second cannot be strongly produced in most CCSN models. Though some rapidly rotating massive stars can produce Ba \citep[e.g.,][]{limongi2018, choplin2018}, they cannot sufficiently produce most heavy $s$-process elements. Low and intermediate-mass AGB star winds also produce the second $s$-process peak elements, though they cannot reach solar [X/Fe] alone and most heavily contribute at $\feh > -1$ \citep{cristallo2011, kobayashi2020}. Rarer $r$-process events, including MRD SN, may also produce some Ba, La, Ce, and Nd at low metallicity \citep[e.g.,][]{nishimura2015}. We find super solar abundances of [Ba/Fe], [La/Fe], [Ce/Fe], and [Nd/Fe], consistent with the low-metallicity GALAH samples for most elements. Notably, Nd appears offset to larger [Nd/Fe] than the GALAH giants of similar metallicity. The abundances trends for these four elements extend from the MINCE abundance tracks, though we find that our lowest metallicity stars ($-1.8 \lesssim \feh \lesssim -2.1$) have [Ce/Fe] and [Nd/Fe] abundances that are $\sim 0.2-0.6$ dex larger than those of MINCE stars at the same metallicity.

Finally, Sm and Eu are produced through the $r$-process, with only 31\% and 6\%, respectively, of the solar abundances produced in the $s$-process \citep{bisterzo2014}. Their $r$-process production sources include rare events such as merging neutron stars, MRD SN, and collapsars \citep[e.g.,][]{wanajo2014, nishimura2015, siegel2019}. The detailed contributions of such nucleosynthetic sources to GCE are not robustly constrained, as few models with extensive nuclear grids exist. The five Sm and one Eu lines in our spectra are weak, with the single Eu line in a strong blend with Ca (see Figure~\ref{fig:spectra}). We detect Sm in all stars, and detect Eu in 56 stars, reporting upper limits for Eu where the line is not detected. We find super-solar [Sm/Fe] abundances, in agreement with the low-metallicity GALAH sample\footnote{The GALAH Nd, Sm, and Eu 2D abundance distributions show banding near [X/Fe]$\approx 0$, potentially indicative of systematic abundance errors in the unflagged GALAH sample.}.  At $\feh \lesssim -1.5$, the [Sm/Fe] abundances of our sample diverge from the MINCE sample, extending to high [Sm/Fe] while the MINCE abundance trend continues to decrease towards solar. This may indicate we are only detecting an upper limit in these low-metallicity stars and that the true [Sm/Fe] is lower. Conversely, our [Eu/Fe] abundances sit near solar, in agreement with the GALAH sample at higher metallicity but substantially lower than the MINCE sample at $\feh < -1.5$. The divergence from the MINCE sample may indicate different Eu abundances in the parent population or systematic abundance differences between subgiant and giant stars. We also observe a larger amount of scatter in our [Eu/Fe] abundances than in that of the MINCE sample. We stress, however, that our Eu abundances are derived from one heavily blended line, and should be interpreted with caution.

In many of the heavy element abundance distributions shown in Figure~\ref{fig:x_fe} there are a few stars with anomalous abundances.
2MASS J06535189+4605517 appears as a strong outlier in Y, Zr, Ba, La, Ce, and Nd, with [X/Fe] of 1.07 (Zr) to 1.88 (La). We confirm through inspection of the spectra that the star is well fit and appears enhanced in many heavy elements relative to its neighbors in [Fe/H].
Similarly, 2MASS J14200701+3953535 is strongly enhanced in Ba, La, Ce, and Nd. These two stars appear to be Ba stars or S stars, heavily enriched in $s$-process material \citep[e.g.,][]{bidelman1951, keenan1954, jorissen2019}. 
Conversely, 2MASS J08090405+0225205 shows an anomalously low [Ba/Fe] of $-0.55$, as well as low Y, Zn, and other light elements including Mg, Ti, and Sc. We discuss these stars in more detail in Appendix~\ref{appendix:weird_guys} and remove the Ba stars from our main intrinsic scatter analysis.

\section{Intrinsic Abundance Scatter}\label{sec:intrin-scatter}

Our primary goal in this work is to measure the intrinsic scatter in the observed heavy element abundances. In Section~\ref{subsec:photon-noise}, we described our estimate of the median photon-noise contribution to the total scatter. In this section we describe our measurement of the total abundance scatter and our determination of the intrinsic scatter remaining after the contribution from photon-noise scatter is removed. We assume that scatter caused by NLTE effects and systematic variations in abundances with $\teff$ and $\logg$ are negligible in our sample. See Section 5.2 of G23 for a more detailed discussion of the impact of NLTE effects.

To measure the total scatter in our low-metallicity stellar population, we fit the [X/Fe] vs. [Fe/H] abundance trends with a linear model and determine the RMS deviations from the predicted abundances. For this calculation, we remove the two Ba stars, which are extreme outliers in the $s$-process elements. The linear model fits a metallicity-dependent, two-parameter trend, such that
\begin{equation}\label{eq:two-param}
    \xfe_{\rm pred} = \xfe_{\rm cen} + Q^{X}(\feh - \feh_{\rm cen}),
\end{equation}
where $Q^{X}$ and $\xfe_{\rm cen}$ are free parameters, $\feh_{\rm cen}$ is the sample median $\feh$, and $\feh$ is the observed Fe abundance. We fit the model to each sample [X/Fe] and simultaneously optimize $Q^{X}$ and $\xfe_{\rm cen}$ with $\chi^2$ minimization implemented through \texttt{scipy.optimize} with the Nelder-Mead method \citep{scipy}. We take the total RMS scatter to be 
\begin{equation}\label{eq:RMS}
    \sigma_{\rm RMS} = \sqrt{\overline{(\xfe - \xfe_{\rm pred})^2}}.
\end{equation}

This describes the overall variation in the data, which is a measure of both photon-noise scatter and intrinsic scatter. We do not adopt the the median photon-noise as $\sigma_{\rm phot}$, but rather estimate the photon-noise about the model (Equation~\ref{eq:two-param}). For this noise estimate, we place each star at it's [X/Fe] predicted by Equation~\ref{eq:two-param}, then perturb this [X/Fe] by a randomly drawn error from a Gaussian distribution with width equal to the star's standard deviation in [X/Fe] (Section~\ref{subsec:photon-noise}). We then calculate the RMS deviation between these perturbed values and the predicted values. For a more robust estimate of the photon noise, we repeat this process 100 times and take the median RMS deviation as the photon-noise scatter about the two-parameter model, $\sigma_{\rm phot, \,2-param}$. We report $\sigma_{\rm phot, \,2-param}$ for all elements in Table~\ref{tab:scatters}.

We find that the $\sphottwo$ is equal to or larger than the $\sphotmed$ for all elements, with differences as large as 0.02 dex (Co). This difference is not surprising, since stars with larger $\sigma_{\rm phot}$ contribute more to the noise-induced scatter. Photon-noise values for the light elements are in agreement with G23, so we do not discuss them in detail. Among the heavy elements, we find that the $\sphottwo$ values range from 0.032 (Zn) to 0.085 (Eu). The photon-noise about the two parameter model is greater than 0.04, our target precision, for all heavy elements but Zn. It remains below 0.06 for all elements but Cu and Eu. 

\begin{deluxetable*}{cccccc}
\tablecaption{Observed Scatter in [X/Fe] Abundances \label{tab:scatters}}
    \tablehead{\colhead{} &  \colhead{$\sigma_{\rm phot,\, med}$} & \colhead{$\sigma_{\rm phot,\,2-param}$}  &  \colhead{$\sigma_{\rm intrin,\,2-param}$}  &  \colhead{$\sigma_{\rm intrin,\,3-param,\,\alpha}$} &  \colhead{$\sigma_{\rm intrin,\,3-param,\,Ba}$}}
    \startdata
Na & 	 0.021 & 	 0.028 & 	 0.156 & 	 0.138 & 	 0.146 \\
Mg & 	 0.022 & 	 0.028 & 	 0.088 & 	 0.057 & 	 0.079 \\
Si & 	 0.017 & 	 0.022 & 	 0.077 & 	 0.064 & 	 0.067 \\
Ca & 	 0.015 & 	 0.018 & 	 0.061 & 	 0.040 & 	 0.054 \\
Sc & 	 0.029 & 	 0.039 & 	 0.095 & 	 0.067 & 	 0.082 \\
Ti & 	 0.018 & 	 0.023 & 	 0.067 & 	 0.056 & 	 0.062 \\
V & 	 0.027 & 	 0.036 & 	 0.105 & 	 0.104 & 	 0.105 \\
Cr & 	 0.012 & 	 0.014 & 	 0.042 & 	 0.040 & 	 0.041 \\
Mn & 	 0.017 & 	 0.021 & 	 0.066 & 	 0.065 & 	 0.065 \\
Co & 	 0.036 & 	 0.056 & 	 0.143 & 	 0.142 & 	 0.144 \\
Ni & 	 0.009 & 	 0.011 & 	 0.056 & 	 0.050 & 	 0.054 \\
Cu & 	 0.055 & 	 0.069 & 	 0.177 & 	 0.177 & 	 0.177 \\
Zn & 	 0.027 & 	 0.032 & 	 0.113 & 	 0.086 & 	 0.101 \\
Sr & 	 0.043 & 	 0.058 & 	 0.171 (0.189) & 	 0.164 (0.180) & 	 0.156 (0.156) \\
Y & 	 0.048 & 	 0.059 & 	 0.204 (0.236) & 	 0.167 (0.199) & 	 0.152 (0.151) \\
Zr & 	 0.041 & 	 0.048 & 	 0.166 (0.178) & 	 0.152 (0.162) & 	 0.141 (0.140) \\
Ba & 	 0.036 & 	 0.050 & 	 0.176 (0.246) & 	 0.149 (0.221) & 	 --- \\
La & 	 0.043 & 	 0.050 & 	 0.184 (0.262) & 	 0.184 (0.261) & 	 0.149 (0.160) \\
Ce & 	 0.046 & 	 0.054 & 	 0.154 (0.225) & 	 0.154 (0.224) & 	 0.140 (0.159) \\
Nd & 	 0.044 & 	 0.056 & 	 0.139 (0.195) & 	 0.139 (0.195) & 	 0.135 (0.155) \\
Sm & 	 0.048 & 	 0.058 & 	 0.163 (0.178) & 	 0.157 (0.175) & 	 0.162 (0.168) \\
Eu & 	 0.071 & 	 0.085 & 	 0.272 & 	 0.202 & 	 0.271 \\
    \enddata
\tablecomments{Values represent the scatter on the 84 low-metallicity stars, excluding the two outlier Ba-stars. For the $s$-process elements, we provide the intrinsic scatter values when the Ba-stars are included in parenthesis. The Ba stars' inclusion does not effect the intrinsic scatter for other elements.}
\end{deluxetable*}

\begin{figure*}[htb!]
    \centering
    \includegraphics[width=1\linewidth]{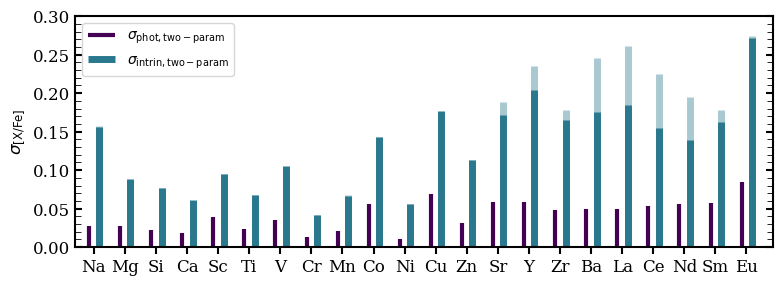}
    \caption{Magnitude of the photon-noise scatter about the two-parameter model (dark purple, $\sphottwo$) and intrinsic scatter about the two-parameter model (blue, $\sigma_{\rm intrin,\,2-param}$) for our sample of [X/Fe] abundances. We show the intrinsic scatter including the two outlying Ba stars as the light blue bars, and the intrinsic scatter with their exclusion as the foreground dark blue bars.}
    \label{fig:scatter}
\end{figure*}

To calculate the intrinsic scatter, we subtract the photon-noise scatter from the RMS scatter in quadrature, such that 
\begin{equation}\label{eq:intrin}
    \sigma_{\rm intrin = \sqrt{\sigma_{\rm RMS}^2 - \sigma_{\rm phot}^2}}.
\end{equation}
We find that the intrinsic scatter about the two-parameter model is greater than 0.1 dex in all elements heavier than Ni, and larger than 0.15 dex for all $s$- and $r$-process elements but Nd.
The Eu scatter is largest, with $\sigma_{\rm intrin, \, two-param}=0.27$.
These measurements are robust, as $\sigma_{\rm intrin, \, two-param}$ is greater than $\sphottwo$ by a factor of $2-4$ for all heavy elements. 
We report the intrinsic scatter about the two-parameter model ($\sigma_{\rm intrin,\, two-param}$) in Table~\ref{tab:scatters}, and plot the intrinsic scatter alongside the photon-noise scatter in Figure~\ref{fig:scatter}. 
In Figure~\ref{fig:scatter}, we see that the intrinsic scatter is notably larger for the heavy elements than for the light elements. We also see that all heavy elements but Zn and Eu have similar intrinsic scatter of $\sim0.17\pm0.03$. 
Notably, the intrinsic scatter amplitudes of light elements Na and Co are large ($\sim0.15$) and more similar to that of the heavy elements than the $\alpha$- and Fe-peak elements. Zn is the only heavy element with $\sigma_{\rm intrin, \, two-param} < 0.13$, appearing more like the $\alpha$ and Fe-peak elements than the neutron capture elements. 
Additionally, we present the intrinsic scatter when the two Ba stars are considered, quoting the values in parenthesis in Table~\ref{tab:scatters} and plotting the intrinsic scatter as the lighter background bars in Figure~\ref{fig:scatter}. When these outliers stars are included, $\sigma_{\rm intrin,\, two-param}$ increases by a factor of $\sim1.4$ for Ba, La, Ce, and Nd. This inflated scatter is not representative of the main low-metallicity sample.

\begin{figure}[htb!]
    \centering
    \includegraphics[width=1\linewidth]{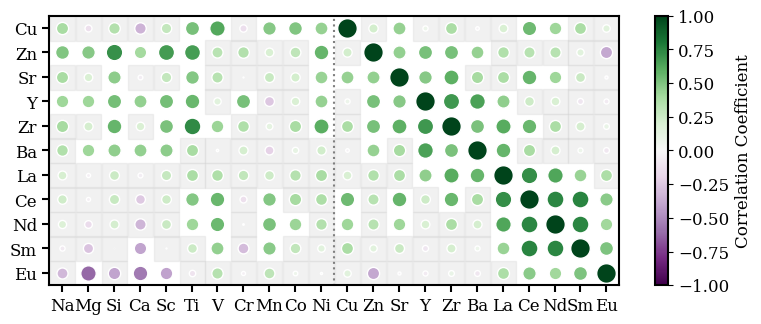}
    \caption{Pearson correlation coefficient ($r$) between the [X/Fe] deviations from the two-parameter model (Equation~\ref{eq:two-param}) for pairs of elements. The size and shade of the circle represents the magnitude of the correlation, with positive values in green and negative values in purple. Correlations between pairs of the same elements are one by definition. Pairs where the correlation coefficient is strong ($|r|>0.4$) are highlighted with a white background, while pairs where the correlation is weaker ($|r|<0.4$) have a gray background.}
    \label{fig:correlation}
\end{figure}

To test if stars with an excess of one abundance have correlated or anti-correlated abundances of other elements, we calculate the the Pearson correlation coefficient ($r$) between the [X/Fe] deviations from the two-parameter model for each pair of elements, excluding the outlying Ba stars. We show these values for the heavy elements in Figure~\ref{fig:correlation}, where the size and color of the circle indicates the correlation strength and sign. The correlation matrix shows a strong correlation ($r\geq0.4$) between many of the $s$-process elements, with most pairs between Sr, Y, Zr, Ba, La having strong correlations. $r$-process elements Sm and Eu are also positively correlated ($r\geq 0.4$) with second peak $s$-process elements La, Ce, and Nd and are positively correlated with each other. 
The observed strong positive correlations indicate that these elements are likely produced in similar sources, with stochastic sampling of the IMF or variations in the source fractions causing higher (or lower) abundances of correlated elements in a given star. Interestingly, the deviations from the two-parameter model for the Fe-cliff elements Cu and Zn are not correlated, though Zn is correlated with many first $s$-process peak elements (e.g., Sr, Y, Zr) and Cu with second $s$-process peak elements Ce and Nd. This suggests that Cu and Zn may have different dominant nucleosynthetic sources at low metallicity. 

Additionally, we find correlations between the heavy elements and light elements. We see a positive correlation ($r\geq 0.4$) between most $\alpha$ elements and Fe-cliff/$s$-process elements Zn, Y, and Ba. This may suggest that at low metallicity the first peak $s$-process elements are created in a nucleosynthetic events that also preferentially produce these $\alpha$-elements relative to Fe (e.g., subsets of CCSN). Heavy elements Ce, Nd, and Sm are positively correlated with Mn and V ($r\geq 0.4$), expected if delayed sources including SNIa and AGB stars have enriched our sample on similar timescales. Finally, we find that Eu is negatively correlated with Mg and Ca ($r\leq -0.4$), potentially due to the strong blend of the Eu line with Ca and imperfect abundance determination.

\subsection{Variations in Ratios of Enrichment Sources}\label{subsec:three-param}

One source of intrinsic scatter in [X/Fe] abundances is the star-to-star variation in the ratio of enrichment sources, such as CCSN, SNIa, and AGB stars. While SNIa and AGB stars typically enrich on a delayed timescale, some more prompt events may contribute to chemical evolution at $\feh<-1$. To account for these source of scatter in our metric, we fit three-parameter models to our abundances. 

To isolate the scatter contribution from the ratio of SNIa/CCSN, we fit a the three-parameter $\alpha$ model:
\begin{equation}\label{eq:three-param}
    \xfe_{\rm pred} = \xfe_{\rm cen} + Q^{X}_{\rm Fe}(\feh - \feh_{\rm cen}) + Q^{X}_{\alpha}\afe,
\end{equation}
where $\xfe_{\rm cen}$, $Q^{X}_{\rm Fe}$, and $Q^{X}_{\rm \alpha}$ are free parameters, $\feh$ is the observed Fe abundance, $\feh_{\rm cen}$ is the median Fe abundances of the sample, and $\afe$ is the median abundance of the Mg, Si, and Ca\footnote{When fitting Mg, Si, or Ca we remove it from the calculation of $\afe$}. The additional dependence on $\afe$ in this three-parameter model mimics the models from \citet{weinberg2019, weinberg2022} and \citet{griffith2019, griffith2024} that describe stellar abundances as the sum of a prompt Mg-like process (CCSN) and delayed Fe-like process (SNIa). 

Similarly, to test if variations in the AGB/CCSN ratio contribute to our observed scatter, we fit the three-parameter Ba model:
\begin{equation}\label{eq:three-param-Ba}
    \xfe_{\rm pred} = \xfe_{\rm cen} + Q^{X}_{\rm Fe}(\feh - \feh_{\rm cen}) + Q^{X}_{\rm Ba}{\rm[Ba/Fe]},
\end{equation}
similar to Equation~\ref{eq:three-param}, but where $Q^{X}_{\rm Ba}$ is a free parameters and [Ba/Fe] is the observed Ba abundance. We take Ba as our representative $s$-process element as it is robustly measured in all of our stars. The addition of a Ba-component mimics the simple model from \citet{griffith2022}, which describes heavy elements Ba and Y as the sum of a Mg-like process (CCSN) and a Ba-like process (AGB).

If star-to-star variation in the CCSN/SNIa ratio significantly contributes to the observed scatter, we expect the intrinsic scatter about the three-parameter $\alpha$ model (Equation~\ref{eq:three-param}) to be less than that about the two-parameter model (Equation~\ref{eq:two-param}). Similarly, if variations in the CCSN/AGB ratio contribute to the scatter, we expect the intrinsic scatter about the three-parameter Ba model (Equation~\ref{eq:three-param-Ba}) to be lower.

We refit the stellar abundances with both three-parameter models, again using \texttt{scipy.optimize} $\chi^2$ minimization with the Nelder-Mean method. We calculate the RMS and intrinsic scatter with Equations~\ref{eq:RMS} and~\ref{eq:intrin}, recalculating $\sigma_{\rm phot,\, three-param}$ as described above for the two-parameter model case. The photon noise about both three-parameter models is nearly identical to $\sigma_{\rm phot,\, two-param}$. We report both three-parameter intrinsic scatters in Table~\ref{tab:scatters}. 

\begin{figure*}[htb!]
    \centering
    \includegraphics[width=1\linewidth]{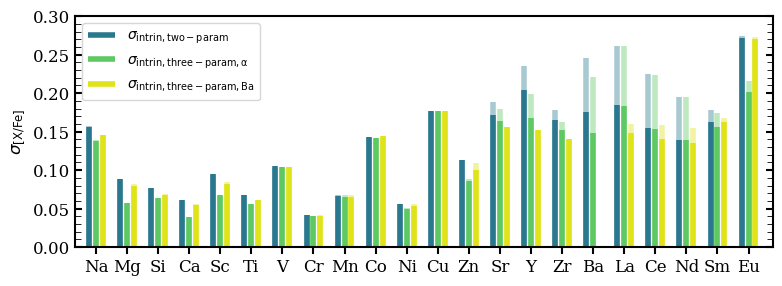}
    \caption{Magnitude of the intrinsic scatter about the two-parameter model (blue, $\sigma_{\rm intrin,\,2-param}$), intrinsic scatter about the three-parameter model with $\alpha$ (green, $\sigma_{\rm intrin,\,3-param,\,\alpha}$) and intrinsic scatter about the three-parameter model with Ba (yellow, $\sigma_{\rm intrin,\,3-param,\,Ba}$) for our sample of [X/Fe] abundances. As in Figure~\ref{fig:scatter}, we show the intrinsic scatters including the outlying Ba stars as the background lighter bars.}
    \label{fig:scatter_three}
\end{figure*}

In Figure~\ref{fig:scatter_three}, we plot the $\sigma_{\rm intrin,\, 3-param,\, \alpha}$ and $\sigma_{\rm intrin,\, 3-param,\, Ba}$ values alongside $\sigma_{\rm intrin,\, 2-param}$. For the three-parameter $\alpha$ model, we see that the intrinsic scatter is notably less than that about the two-parameter model for $\alpha$-elements, Zn, Y, Ba, and Eu. The decrease is a factor of 1.2 (Y) to 1.3 (Zn and Eu). All other elements show a less significant decrease in scatter by a factor of 1.1 or less. This decrease in scatter suggests that star-to-star variations in the ratio of enrichment sources may contribute in a minor way to scatter in Zn, Y, Ba, and Eu. The lower $\sigma_{\rm intrin,\, 3-param,\, \alpha}$ for Eu may indicate that we have not sufficiently de-blended the Eu and Ca line

The three-parameter Ba model also decreases the scatter relative to the two-parameter model for a subset of elements. Heavy elements Zn, Y, Zr, and La see a decrease in scatter by a factor of 1.1 (Zn) to 1.3 (Y), corresponding to 0.01 to 0.05 dex. This suggests AGB star enrichment may make a minor contribution to our sample, but that star-to-star variations in the CCSN/AGB enrichment ratio is not a dominant source of intrinsic abundance scatter. 
When the two Ba stars are included in the calculations of intrinsic scatter, we see a much larger variation in $\sigma_{\rm intrin,\, 3-param, \, Ba}$ and $\sigma_{\rm intrin,\, 2-param}$, as the three parameter Ba model is accounting for the excess $s$-process enrichment in these outlying stars while the two-parameter model is not.
As expected, the scatter about the models does not change for $r$-process elements Sm and Eu.

\section{Discussion: Sources of Intrinsic Scatter}\label{sec:disc}

We have found that the intrinsic abundance scatter about a two-parameter model in [X/Fe] ranges from 0.11 dex to 0.27 dex for the 11 heavy elements ($Z \geq 29$) analyzed in this work. The small photon-noise magnitude ($\lesssim 0.06$ dex) suggests that we are robustly measuring the true intrinsic scatter in this low-metallicity stellar population. The elements studied in this work are produced in a range of nucleosynthetic sources, including CCSN, AGB stars, MRD SN, electron capture SN, collapsars, and neutron star mergers \citep[e.g.,][]{cristallo2011, wanajo2014, nishimura2015, andrews2017, limongi2018, choplin2018, wanajo2018, siegel2019}. In order for the observed intrinsic scatter to arise, the stars studied here must have a range of enrichment histories. This can be caused by mixing of different accreted populations with a diversity of chemical evolution histories, or from non-uniform enrichment across the proto-Galaxy, as stars at different positions are enriched by a finite number of events and un-mixed gas. 

Many mechanisms can lead to abundance scatter, including fluctuations in the ratio of enrichment sources at a fixed metallicity, metallicity or time dependent SN yields, temporal or spatial variations the IMF, and stochastic sampling of the SN population. \citet{belokurov2022} and G23 provide a detailed discussion of how each mechanism contributes to abundance scatter. Here, we focus on stochastic sampling of SN yields and assume that all other scatter sources make minor contributions to the observed intrinsic scatter. This is likely a poor assumption for some cases, such as variations in the ratio of enrichment sources. While we do not expect AGB stars to significantly contribute at $\feh< -1$, the presence of two Ba stars in our sample and slight decrease in the intrinsic scatter about a three-parameter Ba model in Section~\ref{subsec:three-param} suggests that some AGB enrichment has occurred. More prompt AGB star and/or variations in the neutron star merger rate may cause additional star-to-star scatter that is not taken into account with stochastic sampling.

We expect stochastic sampling of the SN population to be a large source of intrinsic abundance scatter at low metallicity. Stochastic GCE models, such as those from \citet{cescutti2013} and \citet{cescutti2014}, produce a spread in abundances similar to that observed by high resolution spectroscopic studies when enrichment events are mixed across regions of $\sim 90$ pc. 
This agrees with the mixing radius estimate of $\sim 100$ pc (or mixing mass estimate of $\sim6\times10^4\msun$) derived in G23, who investigate the impact of stochastic sampling of the IMF on the expected abundance scatter. G23 randomly sample the IMF for $N$ CCSN and calculate the RMS variation in the predicted [X/Fe] distribution for 1000 CCSN samples using yields with a fine mass grid from \citet{sukhbold2016}. They find that stochastic enrichment from roughly 50 CCSN would produce abundance scatter in agreement with the observed intrinsic scatter in the light elements (Na through Ni) for the same stellar population as is studied here. At order-of-magnitude level, if the variation in abundance ratios produced by individual SN is order unity, then the RMS variation from 50 CCSN is $1/\sqrt{50}\approx0.06$ dex. In more detail, the stochastic sampling model gives a reasonably good match to the dependence of scatter on element and to the correlation of star-by-star deviations.

Unfortunately, we cannot extend the theoretical comparison from G23 to the heavy elements, as the yield set from \citet{sukhbold2016} is deficient in producing $s$-process elements, potentially due to uncertainties in neutron producing reaction rates and lack of stellar rotation. Instead, in the following subsections, we explore the impact of sampling stars of different rotational velocities as well as sampling yields from MRD SN. In both cases we assume a parent population of 50 CCSN, corresponding to a mixing mass of $\sim6\times10^4\msun$, following the conclusions of G23. Note that the mixing mass quoted here is the amount of gas required to dilute the yields of $\sim50$ CCSN to a metallicity of $\feh \sim -1.5$, near the middle of our sample.

\subsection{Stochastic Sampling \& Rotating Massive Stars}\label{subsec:rotation}

Rapidly rotating massive stars produce elements in the first $s$-process peak when they explode as CCSN. As suggested by \citet{meynet2006}, rotationally induced mixing increases the amount of $^{14}\rm N$ produced by the CNO cycle, which in turn creates more $^{22}\rm Ne$ \citep{frischknecht2016}. As the $^{22}{\rm Ne}(\alpha,n)^{25}{\rm Mg}$ reaction is a main source of free neutrons needed for the $s$-process, rapid rotation enhances the yields of heavy elements up to Ba. This is seen in theoretical models of rotating massive stars, including \citet{frischknecht2012, frischknecht2016}, \citet{prantzos2018}, \citet{choplin2018}, and \citet{limongi2018}.

Rapid rotation preferentially boosts heavy element yields at low metallicity, as high-metallicity stars tend to rotate slower due to their deeper convective envelopes and increased magnetic activity \citep{amard2020, see2024}, while at lower-metallicity stars are more compact, experience less wind-driven mass loss, and better conserve angular momentum \citep{choplin2018}. Though the rotation rate distribution in the early MW is not well constrained, it is likely that the progenitor massive star population that enriched our low-metallicity stellar sample spanned a range of rotation rates with a large fraction of rapidly rotating stars \citep[e.g.,][for the SMC]{hunter2008}. 

Following the analysis of G23, we model the abundance scatter expected from stochastic sampling of the IMF and the stellar rotation distribution. As stars of different progenitor masses and rotation rates have different yields, the expected yield for a small number ($10-100$) of CCSN events will vary by 0.05 to 0.4 dex over large samples of $N$ CCSN (G23).
Here, we adopt the yields from \citet{choplin2018}. This work calculates $s$-process yields for massive stars at $\feh=-1.8$, providing one set with no rotation and a second set with models rotating at 40\% of the critical velocity---near the peak in the velocity distribution of B-stars observed by \citet{huang2010}. Each set contains nine mass points: 10, 15, 20, 25, 40, 60, 85, 120, 150 $\msun$. We show the individual model yields in Apendix~\ref{appendix:yield}. While in practice some massive stars do not successfully explode and instead collapse to black holes \citep[e.g.,][]{sukhbold2016, ertl2016}, we assume all models successfully explode. The \citet{choplin2018} yield set is not ideal for this analysis, as it has a very sparse mass and velocity grid. However, no yield set that accounts for rapid rotation at low metallicity includes greater than nine mass points. 

\begin{figure*}[htb!]
    \centering
    \includegraphics[width=1\linewidth]{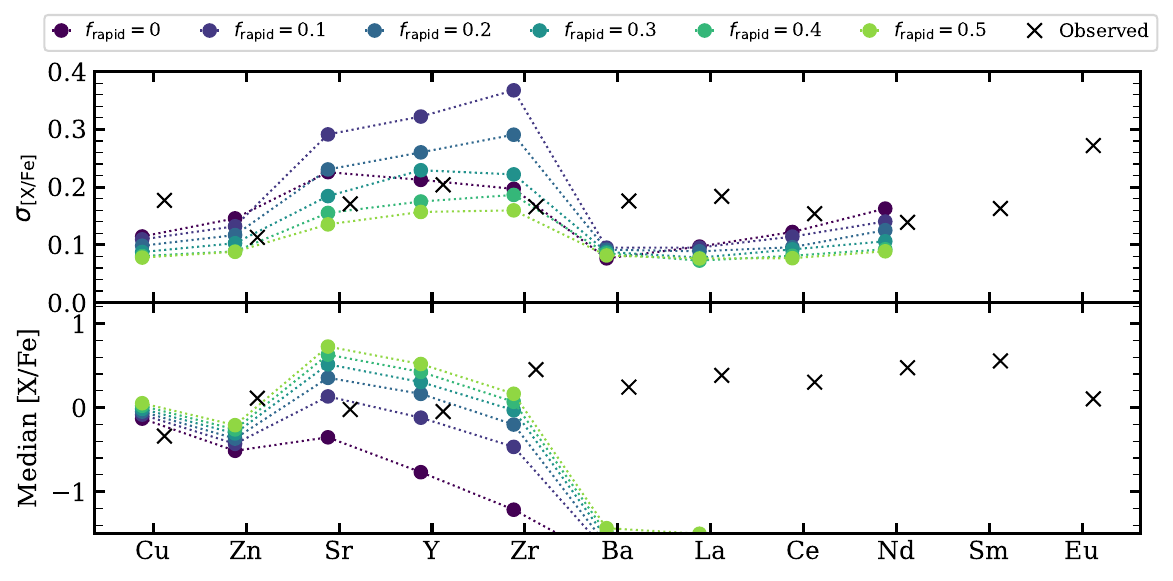}
    \caption{Top: standard deviation of 1000 draws of 50 CCSN with fraction $f_{\rm rapid}$ of stars rotating at 40\% of their critical velocity. We include $f_{\rm rapid}=0.0$ (dark purple), $f_{\rm rapid}=0.1$ (indigo), $f_{\rm rapid}=0.2$ (blue), $f_{\rm rapid}=0.3$ (teal), $f_{\rm rapid}=.4$ (green), and $f_{\rm rapid}=.5$ (lime green). We include the observed intrinsic scatter about the two-parameter model excluding Ba stars ($\sigma_{\rm intrin \, 2-param}$, black x's) for comparison. Bottom: Median [X/Fe] for the same models as above. The median observed [X/Fe] is shown in the black x's. Rotating and non-rotating CCSN yields are from \citet{choplin2018}. Slight horizontal offsets have been added for visual clarity.}
    \label{fig:f_rapid}
\end{figure*}

To estimate the scatter expected from stochastic sampling with varying fractions of rapid rotation, we calculate the standard deviation of the total expected yield in 1000 draws of 50 CCSN with fraction $f_{\rm rapid}$ rotating at 40\% of critical velocity for $f_{\rm rapid}$ = 0, 0.1, 0.2, 0.3, 0.4, and 0.5.
For each draw, we randomly sample 50 stars with masses between $10$ and $150\msun$, the mass range of the \citet{choplin2018} yields, from a \citet{kroupa2001} IMF. Each star is then randomly assigned rapidly rotating or non-rotating yields with probability of rapid rotation $f_{\rm rapid}$. We interpolate the yield in mass space from the appropriate grid and sum the total yields from 50 CCSN. We convert mass yields to [X/Fe] using solar abundances from \citet{asplund2009}. 

In Figure~\ref{fig:f_rapid}, we plot the standard deviation ($\sigma_{\rm [X/Fe]}$, top) and the average [X/Fe] (bottom) of 1000 draws of 50 CCSN with varying $f_{\rm rapid}$ alongside the intrinsic scatter about the two-parameter model and the median observed [X/Fe] abundance of our low-metallicity sample, excluding Ba stars. We find, as expected, that some rapidly rotating CCSN are needed produce the observed abundances of the Fe-cliff and first $s$-process peak elements, but that rapidly rotating CCSN cannot produce sufficient Ba, La, Ce, Nd, Sm, or Eu. The stochastic models with $f_{\rm rapid}$ between 0.2 and 0.4 best match the observed intrinsic scatter, though individual elements are best matched by different fractions of rapid rotation. However, such stochastic models ($f_{\rm rapid}\leq0.4$) cannot produce sufficient quantities of Zn or Zr, suggesting that a larger fraction of rapid rotation is needed or that other nucleosynthetic sources are required. 

We find that the predicted abundance scatter decreases with increasing $f_{\rm rapid}$, for all elements but Y and Zr, where $\sigma_{\rm [Y/Fe]}$ and $\sigma_{\rm [Zr/Fe]}$ increased in draws with $f_{\rm rapid}=0.1$ and 0.2. This is likely due to the fact that the yields for non-rotating and rapidly rotating stars are most different for Y and Zr, especially in the 25 and 40 $\msun$ models (see Appendix~\ref{appendix:yield}). For other elements such as Cu and Zn, the rotating massive star yields are elevated, but more homogeneous than the non-rotating yields, leading to a lower predicted abundance scatter. 
A stochastic model with $f_{\rm rapid}=0$ shows $2-3\times$ the magnitude of scatter as a model with $f_{\rm rapid}=1$. The $f_{\rm rapid}=0$ model severely underpredicts Y and Zr, while the $f_{\rm rapid}=0.5$ model overpredicts Sr and Y. None of these models explains the abundances of elements heavier than Zr. 

Neutrino driven CCSN models with fine mass grids \citep[e.g.,][200 masses]{sukhbold2016} have found that the stellar core compactness, and thus the yield and black hole formation landscape, is complex as a function of progenitor mass. We expect that the predicted scatter magnitude would increase if this analysis were repeated with rotating massive star yields from neutrino driven explosions and a fine mass grid, however, no such yield set currently exists.

\subsection{Stochastic Sampling \& Magnetorotationally Driven Supernovae} \label{subsec:mrd}

Typical CCSN, regardless of rotation rate, cannot produce sufficient quantities of second $s$-process peak elements or $r$-process elements to match observed stellar abundances. While AGB stars may dominate production of $s$-process elements at higher metallicities \citep{cristallo2011}, the predicted AGB yields at $Z<0.1Z_{\odot}$ are drastically lower, so they are unlikely to be a major source of enrichment for our low-metallicity stellar sample. Other, rarer nucleosynthetic events are needed to produce Ba, La, Ce, Nd, Sm, and Eu, with possibilities including MRD SN, electron capture SN, collapsars, and neutron star mergers \citep[e.g.,][]{cescutti2014}. Here, we only investigate the case of MRD SN as a source of these heavy elements. MRD SN, induced by rapid rotation of the stellar core and strong magnetic fields in massive stars, are thought to produce and expel heavy $r$-process material in their jets \citep{symbalisty1984, nishimura2015}. \citet{woosley2006} estimate that, at solar metallicity, 1\% of massive stars explode as MRD SN and that this percentage increases at lower metallicity \citep[e.g.,][]{prochaska2004}. 

To estimate the scatter expected from stochastic sampling of the IMF with a fraction $f_{\rm MRD}$ of massive stars exploding as MRD SN, we conduct a similar estimate as described in Section~\ref{subsec:rotation} but now varying $f_{\rm MRD}$ and including MRD SN yields from \citet{nishimura2015}. \citet{nishimura2015} calculate yields for five $25\msun$ MRD SN models with varying magnetic field and rotation strength. The model with the weakest magnetic field and rotation produces $0.957\times10^{-2}\msun$ of $r$-process material, while the model with the strongest magnetic field and rotation produces a factor of 2.5 more. We note that progenitor the models are all have an initial composition that is approximately solar \citep{heger2000}. See Appendix~\ref{appendix:yield} for a more detailed discussion of the yields.

As the MRD SN rate and yields are uncertain, we estimate the scatter from 1000 draws of 50 CCSN where 2.5\%, 5\%, 7.5\%, 10\%, and 12.5\% of events are MRD SN. For each successful MRD SN, we randomly draw a yield from one of the five models. Note that there is only one mass point in this grid, so all stars contribute the yield of a 25$\msun$ star regardless of their progenitor mass. For the remaining fraction of typical SN, we take CCSN yields from \citet{choplin2018} with $f_{\rm rapid}=0.3$, as described above. We take the standard deviation of 1000 draws of 50 CCSN as the expected scatter, $\sigma_{\rm [X/Fe]}$.

\begin{figure*}[htb!]
    \centering
    \includegraphics[width=1\linewidth]{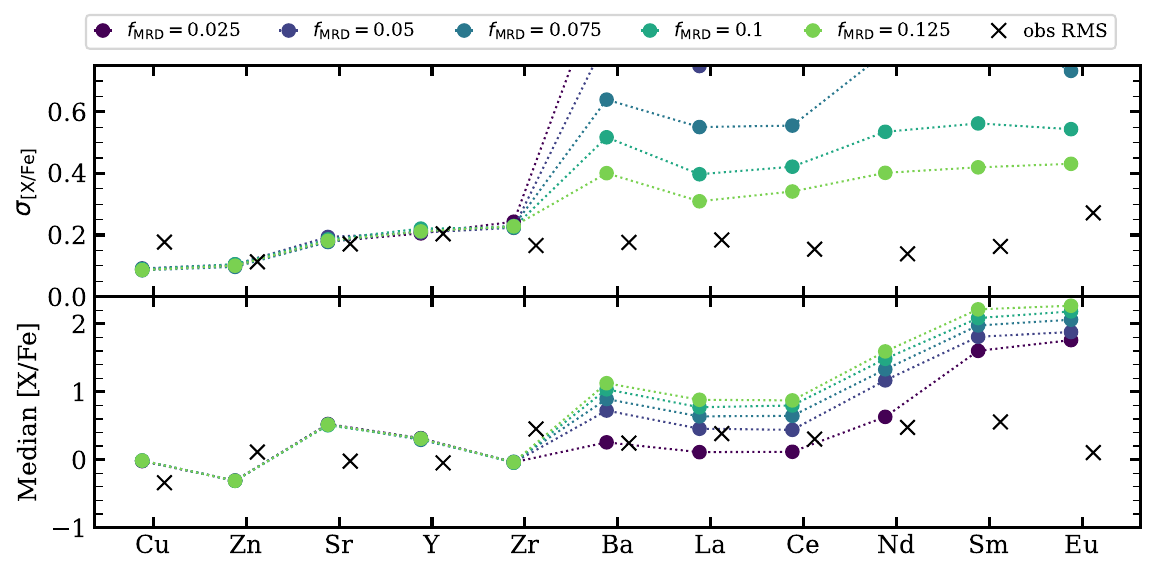}
    \caption{Same as Figure~\ref{fig:f_rapid}, but now for 1000 draws of 50 CCSN with fraction $f_{\rm MRD}$ of stars exploding as MRD SN for $f_{\rm MRD}=0.025$ (dark purple), $f_{\rm MRD}=0.05$ (indigo), $f_{\rm MRD}=0.075$ (blue), $f_{\rm MRD}=0.1$ (teal), and $f_{\rm MRD}=0.125$ (lime green). Yields assume a $f_{\rm rapid}=0.3$. MRD SN yields from \citet{nishimura2015} and CCSN yields from \citet{choplin2018}. Slight horizontal offsets have been added for visual clarity.}
    \label{fig:f_mrsn}
\end{figure*}

We plot the expected scatter and median [X/Fe] from a stochastically sampled IMF with fraction $f_{\rm MRD}$ of events resulting in MRD SN in Figure~\ref{fig:f_mrsn} alongside the observational constraints (excluding Ba stars). While the addition of MRD SN raises the expected [X/Fe] for Ba, La, Ce, Nd, Sm, and Eu to the observed level, all stochastic models result in larger scatter than is observed for the second $s$-process peak and $r$-process elements. The $f_{\rm MRD}=0.125$ model is the closest to observations, but it still exhibits $\sigma_{\rm [X/Fe]}$ significantly larger than the observed intrinsic scatter. While the median observed [X/Fe] for Ba, La, Ce, and Nd is reasonably matched by the stochastic model with $f_{\rm MRD}=0.025$, this low MRD SN fraction significantly over predicts $\sigma_{\rm [X/Fe]}$, reaching values of 1.0 to 1.5 dex. This overprediction is unsurprising. With our estimate of $N\sim50$ CCSN from the light-element scatter, an $f_{\rm MRD}=0.025$ implies only $\sim1.25$ MRD SN enriching a typical star in our sample, so the scatter of enrichment is very large. Notably, all models also over produce Sm and Eu, and the $f_{\rm MRD}=0.125$, the best match to the scatter, significantly overproduces all second $s$-process peak and $r$-process elements. Increasing the number of CCSN sampled in our stochastic models from 50 to 100 brings $\sigma_{\rm [X/Fe]}$ of the $f_{\rm MRD}=0.075$ and 0.1 models into better agreement with the observed intrinsic scatter.

The calculations in Figures~\ref{fig:f_rapid} and~\ref{fig:f_mrsn} are limited by the sparsity of theoretical SN grids at low metallicity, in mass, rotation rate, and magnetic field strength for MRD SN. For the same reason, we have not investigated the predicted correlation of deviations as we did for lighter elements in G23.  Nonetheless, our measured trends and intrinsic scatter provide an empirical target for SN and GCE models in the low-metallicity regime. If the intrinsic scatter does indeed arise from stochastic sampling of the SN population, then the factor $\sim 2$ larger intrinsic scatter for most of the heavy elements suggests that the events that dominate their production are rarer than CCSN but not by a drastic factor.  Although none of the sources considered here explain the observed [Sm/Fe] or [Eu/Fe], the scatter in these ratios is comparable to that of Ba or La, which suggests that the $r$-process events also cannot be drastically more rare.  We caution, however, that roughly $40\%$ of our stars have only upper limits on [Eu/Fe] at our sensitivity.

An alternative to stochastic sampling is that heavy elements come partly from sources with a different delay time distribution from the CCSN that dominate the light elements, and mixing of stellar populations formed with different star formation histories leads to scatter in the ratio of heavy to light elements.  This kind of mixing apparently does account for much of the observed scatter in [$\alpha$/Fe] ratios at the higher metallicities of the MW disk, as demonstrated by the fact that a 2-parameter fit characterizing the ratio of SNIa/CCSN can greatly reduce the abundance ratio scatter for many elements in both the high-Ia and low-Ia populations \citep[e.g.,][]{weinberg2022, mead2025, sit2025}.  However, for the light elements studied by G23 in our low-metallicity sample, adding [$\alpha$/Fe] as a parameter does not remove most of the scatter for individual elements, so this scatter cannot arise mainly from varying the ratio of SNIa/CCSN.  For the $s$-process elements studied here, AGB enrichment by intermediate mass stars would be the natural candidate for an enrichment process that is delayed relative to massive stars. However, the predicted AGB yields at these metallicities are much lower than at solar-like metallicities, so it is unlikely that AGB stars can dominate either the mean or the scatter of [X/Fe] ratios. 

\section{Summary}\label{sec:summary}

In this work we present the abundances of 23 elements (Na, Mg, Si, Ca, Sc, Ti, V, Cr, Mn, Fe, Co, Ni, Cu, Zn, Sr, Y, Zr, Ba, La, Ce, Nd, Sm, and Eu) for 86 low-metallicity ($-2 \lesssim\feh\lesssim-1$) subgiant stars, building off of the initial analysis of the light elements done in G23. Abundances are derived from high-resolution ($R=50,000$) optical spectra taken by PEPSI on the LBT. We adopt stellar parameters ($\teff$, $\logg$, $v_{\rm micro}$) derived with equivalent width analysis in MOOG from G23 and calculate remaining stellar parameters ($v\sin(i)$) and stellar abundances using state of the art spectral synthesis code, \texttt{Korg} (Section~\ref{subsec:synthesis}). Our re-analysis of the light elements ($N\leq28$) is in strong agreement with the values derived in G23 (Figure~\ref{fig:comp_G23}).

We compare heavy element abundance trends derived here with abundances from GALAH DR4 \citep{buder2025} at higher metallicity and from the MINCE survey \citep{cescutti2022,francois2024} at lower metallicity (Section~\ref{subsec:abd_trends}). We generally find good agreement between the three [X/Fe] vs. [Fe/H] trends, though our Ce, Sm, and Eu deviate from the MINCE sample, especially at low metallicity. Visually, we observe more scatter in the heavy elements than was seen for the light elements in G23.

Using multi-parameter linear models, we calculated the total RMS scatter, photon-noise scatter, and intrinsic abundance scatter in our stellar sample for each [X/Fe] in Section~\ref{sec:intrin-scatter} and Figure~\ref{fig:x_fe}. We find large intrinsic scatters in the heavy elements---ranging from 0.11 dex (Zn) to 0.27 dex (Eu), with most heavy element intrinsic scatters near 0.17 dex. While most light elements have smaller scatter ($\sim0.07$ dex), Na and Co appear more like the heavy elements with intrinsic scatter near 0.15 dex. 

The heavy elements studied here have likely been produced by a range of nucleosynthetic sources, with first $s$-process peak elements (Sr, Y, Zr), second $s$-process peak (Ba, La, Ce, Nd), and $r$-process elements (Sm, Eu) forming through different channels. This is highlighted in our observations by correlated [X/Fe] residuals from the best fit linear, two-parameter model, shown in Figure~\ref{fig:correlation}. Here, first and second $s$-process peak elements have strongly correlated residuals as well as the second $s$-process peak and $r$-process elements. This highlights that $r$-process events likely also formed La, Ce, and Nd, but are not the predominant nucleosynthetic source of the first $s$-process peak elements. While SNIa and AGB stars dominate Fe-peak and $s$-process yields at higher metallicity, they do not appear to heavily contribute to abundance scatter within this low-metallicity sample.

Intrinsic abundance scatter can be a tool to constrain early Galactic enrichment and properties of low-metallicity nucleosynthesis. In Section~\ref{sec:disc}, we leverage our precise measurements of intrinsic scatter in the low-metallicity regime to constrain stochastic properties of SN---specifically the fraction of rapid rotation and fraction of MRD SN needed to match the absolute abundances and intrinsic abundance scatters. 
For this analysis, we build a model of stochastic enrichment and measure the scatter in 1000 samples of 50 CCSN using rapidly rotating CCSN yields from \citet{choplin2018} and MRD SN yields from \citet{nishimura2015}.
We find that the scatter in the first $s$-peak elements is best fit by a model with fraction of rapidly rotating CCSN $f_{\rm rapid}$ between 0.2 to 0.4, though a higher fraction of rapid rotation is needed to reach the observed median [Zn/Fe] and [Zr/Fe] abundances.
An $r$-process source, such as MRD SN, is needed to produce the observed abundances of the heaviest elements. Based on the rate estimates from \citet{woosley2006}, we explore stochastic models with $f_{\rm MRD}$ ranging from 0.025 to 0.125. While the intrinsic scatter is best matched by a model with $f_{\rm MRD}=.125$, this model significantly overproduces Nd, Sm, and Eu. 

The analysis conducted here highlights the need for more comprehensive yield sets for SN at low, but non-zero, metallicity. Comparing predicted and observed scatter could be a powerful predictor of early Galaxy and dwarf galaxy nucleosynthesis. However, sufficient yield sets accounting for the complex black hole formation landscape, rotation rates, magnetic fields, and other properties do not exist. Without fine model grids we cannot probe the expected stochastic signature of these events at low metallicity.

\section*{Acknowledgments}   

We thank Adam Wheeler for his guidance on how to best use Korg for our abundance analysis. We also thank Adam Leroy, Tuguldur Sukhbold, Yuan-Sen Ting, Todd Thompson, Fiorenzo Vincenzo, James Johnson, the Ohio State University Galaxy Group, and the Darling research group at CU Boulder for helpful discussions of abundance scatter, nucleosynthetic yields, and stochastic sampling of the IMF. 
We thank the many LBT observers who collected data for this paper: S. Bose, N. Emami, P. Garnavich, R.T. Gatto, C. Howk, T. Jayasinghe, C. Kochanek, O. Kuhn, A. Pai Asnodkar, D. Reed, B. Rothberg, J. Sullivan, A. Tuli, P. Vallely, M. Whittle, J. Williams, and C. Wood.

E.J.G. acknowledges support for this work provided by NASA through the NASA Hubble Fellowship Program grant No. HST-HF2-51576.001-A awarded by the Space Telescope Science Institute, which is operated by the Association of Universities for Research in Astronomy, Inc., for NASA, under the contract NAS 5-26555. E.J.G. was supported during part of this work by an NSF Astronomy and Astrophysics Postdoctoral Fellowship under award AST-2202135. M.B. acknowledges support from the University of Colorado Undergraduate Research Opportunity Program.

The LBT is an international collaboration among institutions in the United States, Italy, and Germany. LBT Corporation partners are as follows: The University of Arizona on behalf of the Arizona Board of Regents; Istituto Nazionale di Astrofisica, Italy; LBT Beteiligungsgesellschaft, Germany, representing the Max-Planck Society, The Leibniz Institute for Astrophysics Potsdam, and Heidelberg University; The Ohio State University, representing OSU, University of Notre Dame, University of Minnesota and University of Virginia.

\software{
Astropy \citep{astropy2013, astropy2018, astropy2022},
\texttt{Korg} \citep{wheeler2023,wheeler2024},
Matplotlib \citep{hunter2007}, 
NumPy \citep{harris2020}, 
Pandas \citep{pandasa, pandasb}, 
PyJulia \citep{pyjulia}
}

\appendix

\section{Outlier Stars}\label{appendix:weird_guys}

\begin{figure*}[htb!]
    \centering
    \includegraphics[width=1\linewidth]{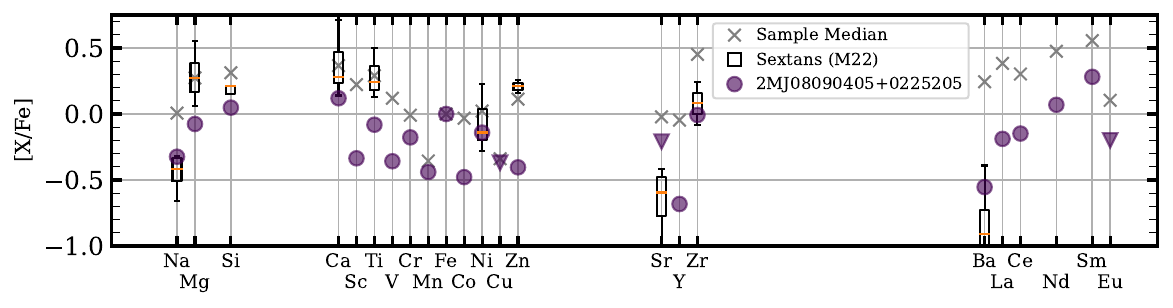}
    \caption{Abundances for 2MASS J08090405+0225205 (dark purple circles). Abundances for which we only detect an upper limit are marked with an upside down triangle. We plot the median sample abundances from this work as the light grey x's and the show the distribution of abundances for stars in the Sextans galaxy from \citet{mashonkina2022} as the black box plots.}
    \label{fig:weird_star}
\end{figure*}

While most stars in our sample follow the population trend, three stars are notable outliers. 2MASS J08090405+0225205 is an outlier in many elements, notably displaying low [X/Fe] for Ba and Y in Figure~\ref{fig:x_fe}, as well as many light elements in G23. We plot the stars abundances vs. atomic numbers in Figure~\ref{fig:weird_star}. Here it is clear that this star has a unique abundances. The star show an odd-even pattern from Na through Ni, where elements with odd atomic numbers are more depleted than elements with even atomic numbers, though all light elements are depleted relative to the sample median. The star also has low abundances of [Zn/Fe], [Y/Fe], and [Ba/Fe]. We find a trend of increasing abundance of the second $s$-process peak elements from subsolar Ba, La, and Ce to super solar Nd and Sm. The significantly subsolar [Ba/Fe] and [Na/Fe] are commonly seen in classical dwarf spheroidal galaxies. In Figure~\ref{fig:weird_star}, we plot the distribution of abundances in the Sextans galaxy from \citet{mashonkina2022}, showing the first and third inter-quartile range and the median in orange. While this dwarf spheroidal shows similarly low abundances of [Na/Fe], [Sr/Fe], and [Ba/Fe], the $\alpha$-elements are super solar compared to the solar abundances seen in 2MASS J08090405+0225205. No abundance pattern from the dwarf spheroidals represented in the JINAbase \citep{abohalima2018} match this the abundance pattern observed for this star. We check the star's kinematics and find that it is consistent with the edge of the thick disk, placing it in a parameter space that is spanned by both in-situ and accreted populations. 

\begin{figure*}[htb!]
    \centering
    \includegraphics[width=1\linewidth]{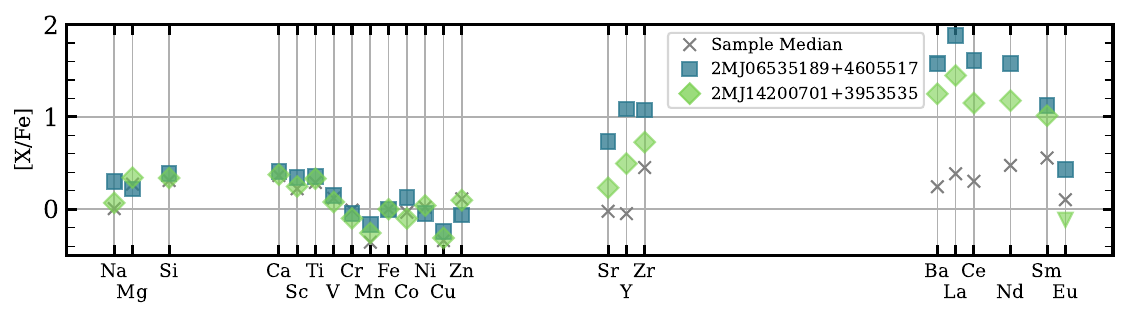}
    \caption{Abundances for 2MASS J06535189+4605517 (blue squares) and 2MASS J14200701+3953535 (green diamonds). Abundances for which we only detect an upper limit are marked with an upside down triangle. We plot the median sample abundances from this work as the light grey x's.}
    \label{fig:ba_star}
\end{figure*}

We also identify two stars with very enhanced Ba and other $s$-process elements in Section~\ref{subsec:abd_trends}. We show the abundances of 2MASS J06535189+4605517 and 2MASS J14200701+3953535 in Figure~\ref{fig:ba_star} alongside our sample medians. This figure clearly shows that while the stars have typical abundances of the lighter elements, they are enhanced in Sr through Sm, with abundance of Ba, La, and Ce that are roughly 1 dex higher than the sample median. These stars are likely Ba stars \citep{bidelman1951} or S stars \citep{keenan1954}. As subgiant stars cannot synthesize $s$-process material, they have likely experienced mass transfer of material rich in $s$-process elements from an AGB binary companion \citep{mcclure1980, jorissen1993}. Many known Ba and S stars have had white dwarf binary companions, the remnant of the past AGB star, identified \citep[e.g.,][]{jorissen2019}. We note that both stars are best fit by $v\sin (i) < 5$ km/s, indicative that neither are rapidly rotating---a signature of binarity and past mass transfer \citep[e.g.,][]{patton2024}. Additional observations and analysis would be needed to further classify these stars and search for the presence of companion remnant.

\section{Theoretical Yields}\label{appendix:yield}

\begin{figure*}[htb!]
    \centering
    \includegraphics[width=1\linewidth]{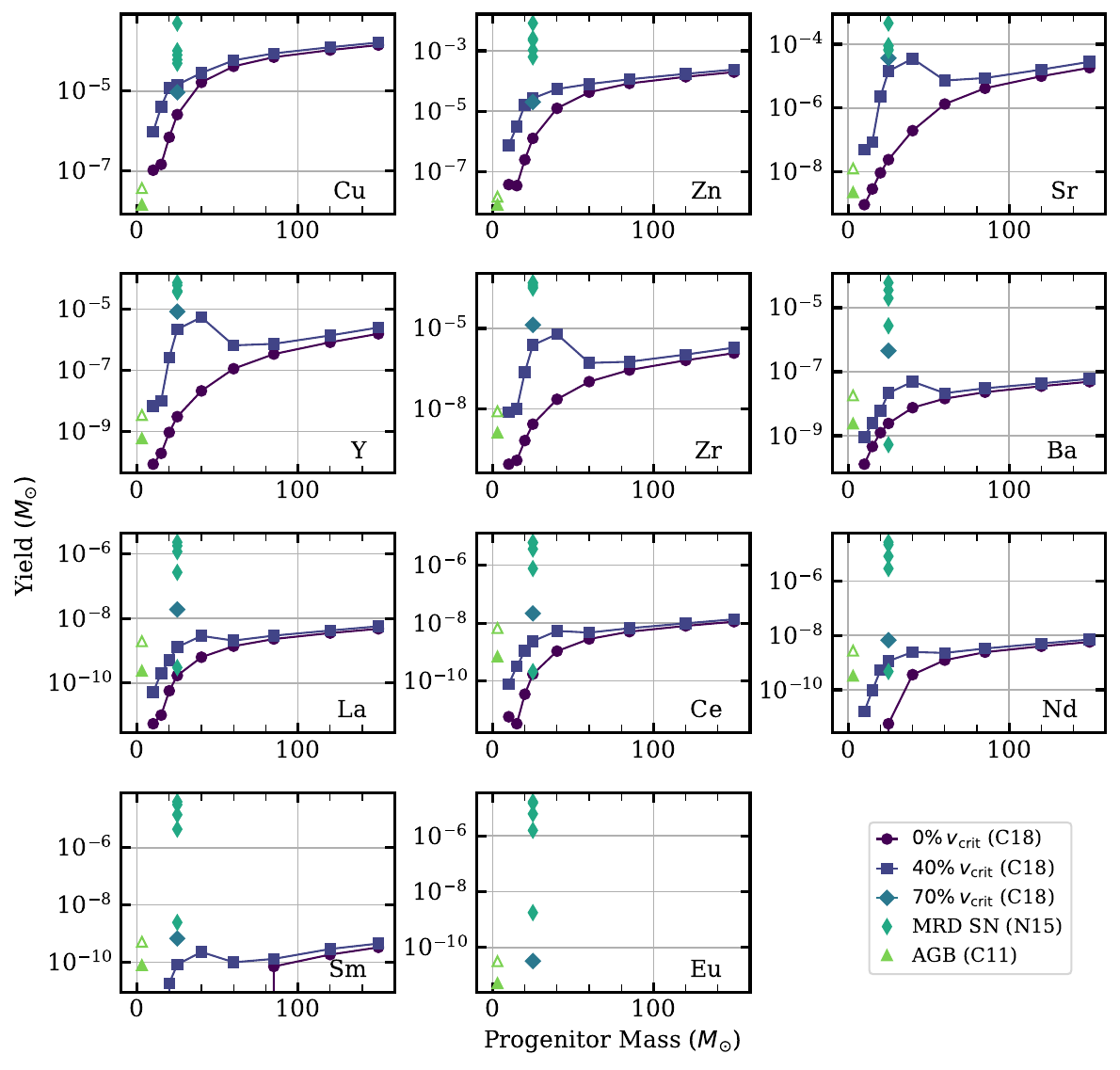}
    \caption{\citet{choplin2018} CCSN yields at $Z = 10^{-3}$ as a function of progenitor mass for the non-rotating yield set (purple circles) and the set rotating at 40\% of critical velocity (dark blue squares) as well as the single $25\msun$ model rotating at 70\% critical velocity (blue wide diamond). We additionally show the \citet{nishimura2015} yields for five solar metallicity MRD SN of with progenitor mass of $25 \msun$ (teal diamonds) and AGB yields for a $2.5\msun$ star with $Z=10^{-4}$ (filled green triangle) and $Z=10^{-3}$ (open green triangle).}
    \label{fig:C18_yields}
\end{figure*}

In Figure~\ref{fig:C18_yields}, we plot the yield of heavy elements in $\msun$ for rotating CCSN from \citet{choplin2018}, MRD SN yields from \citep{nishimura2015}, and AGB star yields from \citet{cristallo2011}. For the \citet{choplin2018} yields, we show both the models with zero rotation, 40\% of critical velocity, and 70\% of critical velocity (only available for $M=25\msun$). All models are at $Z = 10^{-3}$. We see that the production of all elements is boosted in the rotating models, but that elements heavier than Zr are not heavily produced in any model. Elements including Sr, Y, Zr, and Ba see a large boost in production for models between 20 and 40 $\msun$ due to the presence of extra $^{22}\rm Ne$. See Section 3.2 of \citet{choplin2018} for a discussion of this feature. We also include the yields from the five solar metallicity MRD SN from \citep{nishimura2015} for a $25\msun$ star. The MRD SN yields clearly dominate the production of elements Ba through Eu, though the individual models span a large parameter space. Finally, we show the yields for a $2.5\msun$ AGB star at $Z=10^{-4}$ (filled) and $Z=10^{-3}$ (open). For most elements, the AGB yields are significantly smaller than the yields from \citet{choplin2018}.

\bibliography{sample631}{}
\bibliographystyle{aasjournalv7}

\end{document}